\documentclass[]{article}

\usepackage[T1]{fontenc}
\usepackage[latin1]{inputenc}
\usepackage{amssymb}
\usepackage{amsmath}
\usepackage{graphicx}
\usepackage[caption=false]{subfig}
\usepackage{multitoc}
\usepackage{wrapfig}
\usepackage{hyperref}

\usepackage{xcolor}

\usepackage{epstopdf}
\DeclareGraphicsExtensions{.eps, .png}

\usepackage{natbib}

\begin{document}

\title{Impact of Non-potential Coronal Boundary Conditions on Solar Wind Prediction}

%
%

\author{M. Weinzierl\footnote{Department of Mathematical Sciences, Durham University, UK} and F.-X. Bocquet\footnote{Met Office, Exeter, UK} and A.R. Yeates\footnote{Department of Mathematical Sciences, Durham University, UK}}


\maketitle

%
%


\begin{abstract}
Predictions of the solar wind at Earth are a central aspect of space weather prediction. The outcome of such a prediction, however, is highly sensitive to the method used for computing the magnetic field in the corona. We analyze the impact of replacing the potential field coronal boundary conditions, as used in operational space weather prediction tools, by non-potential conditions. For this, we compare the predicted solar wind plasma parameters with observations at 1 AU for two six-months intervals, one at solar maximum and one in the descending phase of the current cycle. As a baseline, we compare with the operational Wang-Sheeley-Arge model.
We find that for solar maximum, the non-potential coronal model and an adapted solar wind speed formula lead to the best solar wind predictions in a statistical sense. For the descending phase, the potential coronal model performs best. The Wang-Sheeley-Arge model outperforms the others in predicting high speed enhancements and streamer interactions. A better parameter fitting for the adapted wind speed formula is expected to improve the performance of the non-potential model here.
\end{abstract}

%
%

\section{Introduction}

\noindent Finding suitable coronal boundary conditions for simulations of the inner heliosphere is a crucial point in solar wind prediction. While a magnetohydrodynamics (MHD) simulation in the coronal domain (like in the MHD-Around-a-Sphere, MAS, model \citep{mikic99mas, lionello2001mas, riley2003mas, riley2001dchb}) would provide accurate boundary data, it is, on today's computers, too time-consuming for use in operational space weather forecasting.
There exist, however, a number of simplified coronal simulation methods, which use, to different degrees, extrapolation and simplifications. Most notably there are the Potential Field Source Surface (PFSS) method \citep{Altschuler69pfss, schatten69pfss}, the magnetofrictional (MF) method \citep{yang86mf, ballegooije00mfmethod, yeates08globalsolarcorona2}, and the Current Sheet Source Surface (CSSS) method \citep{zhao95csss, poduval2014csss, poduval2016csss}. For the outer corona, the Schatten Current Sheet (SCS) method \citep{schatten71scs} is commonly employed. An empirical wind speed formula is then used to compute the boundary conditions for the heliospheric simulation from the coronal simulation data at the interface between the two domains. 
Other methods, for example \citet{reiss2016highspeedstreamforecasts, owens2017similarday, riley2017patternrecognition}, do not simulate the corona (and heliosphere) at all, but rely on purely empirical methods to forecast the solar wind.  Recent research by \citep{pinto2017multivp} proposes an alternative approach as compromise between full MHD simulations and semi-empirical methods. They compute the structure of the solar winds and its parameters by combining a large number of one-dimensional wind profiles along open magnetic field lines. 

\noindent The form of and the parameters for the empirical wind speed formula are a field of research in their own right, although strongly tied to the coronal simulation in use. Common forms are the Wang-Sheeley (WS) model \citep{wang90wsa, wang92wsa, arge2000wsa}, the Wang-Sheeley-Arge (WSA) model \citep{arge2003wsa} and the Distance from the Coronal Hole Boundary (DCHB) model \citep{riley2001dchb}. While the DCHB model has a physics-base explanation of its parameters (which still are varied), the parameters in the WS and WSA formulas are free and adapted for every period considered. \citet{riley2015expansionfactor} have attempted to find optimal parameters for WS, WSA and DCHB by exhaustive search, using a simplified method method in the heliospheric domain. \citet{poduval2014csss, poduval2016csss} determine the coefficients for their wind speed prediction, which is based on the WS formula, by fitting parameters to a quadratic function using observations that were mapped back to the source surface of their simulation.

\noindent Over the last years, there have been a number of attempts to compare and validate the different approaches for predicting the solar wind speed.

\noindent Comparisons of solar wind data of potential and magnetohydrodynamic (MHD) simulations in \citet{riley2006comparemhdpfss} show that, if time-dependent effects can be neglected, the potential method provides a reasonable approximation to the MHD method, although there still are notable differences. 

\noindent \citet{edwards15influencenp} have compared the magnetic structure and the resulting wind speed distribution from potential and non-potential simulations at 21.5 $R_\odot$ for two solar maximum dates. They have identified considerable differences between the two types of coronal simulations: The non-potential model has more complex magnetic structures, more open flux and, using the WSA wind speed formula, leads to higher predicted wind speeds for the two dates. The present work extends this by considering two time intervals of six months each at solar maximum and the descending phase of the solar cycle, and continuing the simulation to 1 AU for comparison with observational data. 

\noindent \citet{jian2015ccmcmodelvalidation, jian2016ccmcmodelvalidation} present an extensive evaluation of a number of coronal and heliospheric models that are available at the Community Coordinated Modeling Centre. They find that there is not a single candidate that performs best, but that each model has its strengths and weaknesses.

\noindent In this work we aim to contribute to the efforts of bringing an order into the multitude of solar wind prediction methods by comparing the impact of the PFSS, MF and WSA method for computing the coronal boundary conditions for solar wind simulations. To investigate the difference between potential and non-potential boundary conditions we distinguish here between WSA and PFSS: We perform the PFSS coronal simulation with the same input data and the same method as the MF one (that is, Air Force Data Assimilative Photospheric Flux Transport (ADAPT) synoptic $B_r$ maps), except that we compute the potential field solution in inner corona. The WSA run, in contrast, uses the National Oceanic and Atmospheric Administration / National Weather Service Space Weather Prediction Center (NOAA/NSW SWPC) operational method. This is also based on a PFSS model for the inner corona, but using observational magnetograms from the National Solar Observatory (NSO) Global Oscillation Network Group \citep[GONG;][]{harvey96gong} directly as input. We pick one six-month interval at solar maximum and one in the descending phase of the solar cycle for comparison and do both a statistical and an event-based comparison.

\noindent The remainder of the paper is structured as follows: In Section \ref{sec:data_methods} we describe the data and the simulation methods that we use in this work. Section \ref{sec:models} details the solar wind models that are used for the different runs. We present our results in Section \ref{sec:results} and conclude and give an outlook on future work in Section \ref{sec:conclusion}.


\section{Data and Methods \label{sec:data_methods}}

\noindent The models were set up as follows (see also Table \ref{tab:models}): WSA model driven by daily updated GONG magnetograms; PFSS which used the DuMFriC PFSS model \footnote{https://github.com/antyeates1983/pfss} driven by ADAPT magnetograms; MF model which used the DuMFriC non-potential (NP) MF model driven by ADAPT magnetograms.

\begin{table}
\caption{Description of the solar wind models \label{tab:models}}
\centering
\begin{tabular}{l | r r r r r}
\hline
 Model name & data source & coronal model & $v_r$ formula  & outer model\\
\hline
  WSA  & GONG &	WSA & Equation (\ref{eq:WSA}) & Enlil    \\
  PFSS & ADAPT & DuMFriC PFSS & Equation (\ref{eq:windspeed}) & Enlil  \\
  MF   & ADAPT & DuMFriC NP & Equation (\ref{eq:windspeed}) & Enlil  \\
\hline
\multicolumn{5}{l}{}
\end{tabular}
\end{table}

\noindent The time intervals chosen for simulation are May 1 to October 31 in 
2014 (solar maximum, CR 2149 to 2156) and 2016 (descending phase, CR 2176 to 2183) respectively. We chose these times for their position in the solar activity cycle and due to data availability.

\noindent All Enlil and OMNI solar wind data is transformed to the Heliocentric Earth Equatorial (HEEQ) coordinate system for the comparisons.

\subsection{ADAPT Dataset}

\noindent The photospheric boundary conditions for the coronal simulation are derived from Air Force Data Assimilative Photospheric Flux Transport (ADAPT) synoptic $B_r$ maps \citep{arge10adapt, henney12adapt, hickmann15adaptadvancements}. The ADAPT maps are constructed from GONG magnetograms by evolving them using a photospheric flux transport model which is based on the Worden-Harvey model \citep{wordenharvey00magneticfluxmaps}. New data are assimilated into the model once per day (weather permitting) and maps are output at twelve hour cadence. In the current version, the ADAPT data set consists of an ensemble of twelve realizations which account for model parameter uncertainties in the supergranular flow. We picked realization number one for our experiments. This choice is expected to have only a minimal influence on the results (see \citep{weinzierl16photosphericdriving}).

\subsection{Coronal and Inner Heliospheric Simulation \label{sec:coronalsim}}

\noindent  For the Potential Field Source Surface (PFSS) model \citep{Altschuler69pfss, schatten69pfss}, the magnetic field between 1 $R_\odot$ and 2.5 $R_\odot$, i.e., in the inner corona, is computed by extrapolation from the photospheric magnetic field, assuming that the field is is current free ($\nabla \times \mathbf{B} = 0$) and radial at 2.5 $R_\odot$.

\noindent The non-potential coronal model simulates the evolution of the large-scale magnetic field  between 1 $R_\odot$ and 2.5 $R_\odot$ using the magneto-frictional (MF) method \citep{yang86mf, ballegooije00mfmethod, yeates08globalsolarcorona2}. Here, the velocity $\mathbf{v}$ is approximated by the magneto-frictional form: $\mathbf{v} = \nu^{-1} (\mathbf{J} \times \mathbf{B}/B^2)$, where $\mathbf{J} = \nabla \times \mathbf{B}$ and $\nu$ is a friction coefficient. This enforces the relaxation of the magnetic field towards a nonlinear force-free state where $\mathbf{J}\times\mathbf{B}=0$. The MF model allows for a gradual build-up and conservation of magnetic energy and electric currents in the corona. 

\noindent The temporal evolution of $\mathbf{B} = \nabla \times \mathbf{A}$ is driven by photospheric $B_r$ maps from which the update $\partial\mathbf{A}/\partial t = - \mathbf{E}$ for the vector potential $\mathbf{A}$ is computed. The method used for the electric field reconstruction is described in \citep{weinzierl16photosphericdriving} and based on work by \citet{amari03cme, fisher10efields, kazachenko14efields}.

\noindent The coronal simulation uses a grid that is equally spaced in $\rho$, $s$, $\phi$, where $\rho = \ln\left(r/R_\odot\right)$ and $s=\cos\theta$ in terms of spherical coordinates $(r,\theta\,\phi)$. The resolution is $60 \times 180 \times 360$. 

\noindent For extrapolation of the magnetic field in the outer corona (2.5 $R_\odot$ to 21.5 $R_\odot$) we use the Schatten Current Sheet (SCS) method \citep{schatten71scs}. Here, we solve for a potential field using the absolute values of the radial magnetic field component at $r = 2.5 R_\odot$ and the assumption that $\mathbf{B} \xrightarrow[r \rightarrow \infty]{} 0$. Then, the field line direction is restored where $B_r<0$ at 2.5 $R_\odot$, producing infinitesimally thin current sheets.

\noindent The magnetic field at 21.5 $R_\odot$ and the expansion factor and coronal hole boundary distance as described in Sec.~\ref{sec:models} are used to compute the boundary conditions for the solar wind software Enlil \citep{odstrcil96enlil, odstrcil03enlil}, which simulates the solar wind in the heliosphere. In this work, we are interested in the wind speed, proton density and magnetic field at Earth. We run Enlil with low resolution, which means 256 cells in $r$, 30 cells in $\theta$ and 90 cells in $\phi$.

\section{Solar Wind Models \label{sec:models}}

\noindent As described in the introduction, most solar wind models consist of two parts. An inner model covering the domain from $R_{\odot}$ to $21.5R_{\odot}$ and an outer model covering the domain from $21.5R_{\odot}$ to some outer boundary such as 1.7AU. The following sections detail the inner models used in this study to determine the boundary conditions for the outer model. The baseline model is the WSA model which is used operationally worldwide. In this paper, we introduce a new approach to modelling the coronal part of the solar wind modelling chain in Section \ref{sec:DUMFRICmethod}, consisting of a potential or non-potential reconstruction of the coronal field using the DuMFriC code driven by ADAPT maps feeding into a alternative empirical formulation for the solar wind speed based on the DCHB model by \citet{riley2001dchb}. This provides alternative boundary conditions to those obtained from the WSA model.

\subsection{Wang-Sheeley-Arge Model \label{sec:WSAmethod}}

\noindent The empirical solar wind model used in operational space weather forecasting at the NOAA/NSW SWPC is the Wang-Sheeley-Arge (WSA) model \citep{arge2003wsa}. It computes the wind speed $v_r$ from the flux tube expansion factor $f_s = \left(\frac{R_\odot}{R_{s}}\right)^2 \left(\frac{B_r(R_\odot)}{B_r(R_s)}\right)$, with $R_s = 2.5 R_\odot$ in our case, and $\theta_b$ (in degrees), which is the minimum distance of a fieldline footpoint from a coronal hole boundary in the photosphere. The WSA formula reads

\begin{equation}
v_r(f_s, \theta_b) = v_{slow} + \frac{v_{fast} - v_{slow}}{(1 + f_s)^{\alpha}} \left[ \beta - \gamma e^{-(\theta_b/\omega)^\delta }\right]^{\iota}. 
\end{equation}

There are eight free parameters, including the fast and slow wind speed $v_{fast}$ and $v_{slow}$. 

\noindent The WSA formula was developed from the Wang-Sheeley (WS) model \citep{wang90wsa, wang92wsa, arge2000wsa}, which only uses $f_s$ as an input parameter and reads 

\begin{equation} \label{eq:WSA}
v_r(f_s) = v_{slow} + \frac{v_{fast} - v_{slow}}{(f_s)^\alpha},
\end{equation}

and the Distance from the Coronal Hole Boundary (DCHB) formula \citep{riley2001dchb} which only uses $\theta_b$ and reads 

\begin{equation}
v_r(\theta_b) = v_{slow} + (v_{fast} - v_{slow})\left[1 + \tanh\left(\frac{\theta_b - \epsilon}{w}\right)  \right], 
\end{equation}

with $\epsilon$ the thickness of the slow flow band and $w$ the width over which flow is raised to coronal hole values. Recent research suggests that $\theta_b$ is more important than $f_s$ for determining the wind speed, and in some cases the presence of $f_s$ actually weakens the predictive power of the WSA formula \citep{riley2015expansionfactor}. However, \citet{mcgregor2011vrsolarmin} found evidence that the expansion factor might influence the distribution of fast solar wind deep inside coronal holes, and that both factors ($\theta_b$ and $f_s$) might be important to accurately model the solar wind.

\subsection{Modified DCHB Model \label{sec:DUMFRICmethod}}
\noindent All three formulas above were developed for input data using a relatively coarse grid, not resolving much of the fine-scale structure of the magnetic field and small (or thin) coronal holes. 
The derived solar wind speed and density maps in \citet{edwards15influencenp} used Equation \ref{eq:WSA} without modifying the parameters used in the WSA model. Here, for our MF and PFSS simulation, we devised a new formula, tentatively cutting down on free parameters and leaving out the expansion factor term. Our tentative empirical wind speed formula reads

\begin{equation} \label{eq:windspeed}
v_r(\theta_b) = v_{slow} + (v_{fast} - v_{slow})(\theta_b \cdot \omega)^\delta,
\end{equation}

with $\theta_b$ in radians, $\omega > 1$, $\delta < 1$, $v_{slow} \in \left[100, 400\right]$ and $v_{fast} \in \left[500, 1000\right]$. Note that if we measured $\theta_b$ in degrees we would need $\omega < 1$.

\noindent Because we use the same high resolution input and coronal hole detection method for the PFSS method we use the same wind speed formula. 
As a first rough parameter fit we use $v_{slow} = 200$ $km s^{-1}$, $v_{fast} = 700$ $km$ $s^{-1}$, $\omega = 7$ and $\delta = 1/2.5$. 

\noindent We find overall that the optimal form of the wind speed formula and the optimal parameter fit are very sensitive to the resolution of the photospheric magnetograms or synoptic maps, the resolution of the coronal simulation and the sensitivity of the coronal hole detection algorithm.


\section{Results \label{sec:results}} 

\noindent In order to compare the solar wind predictions at L1 provided by Enlil when it is driven by the different models, we first produce the boundary conditions separately by the MF, PFSS and WSA model as described in the previous sections. For the MF model, we run DuMFriC with a four-month ramp-up time, i.e., we start the simulation in January of the respective year, in order to make sure that it has reached a valid state by May. For every day, we then do a separate Enlil run with the new boundary conditions derived from the current observational data. 

\subsection{Validation Metrics for Continuous Variables}\label{sec:val:continuous}

\begin{figure}
{\centering
\includegraphics[width=0.45\textwidth]{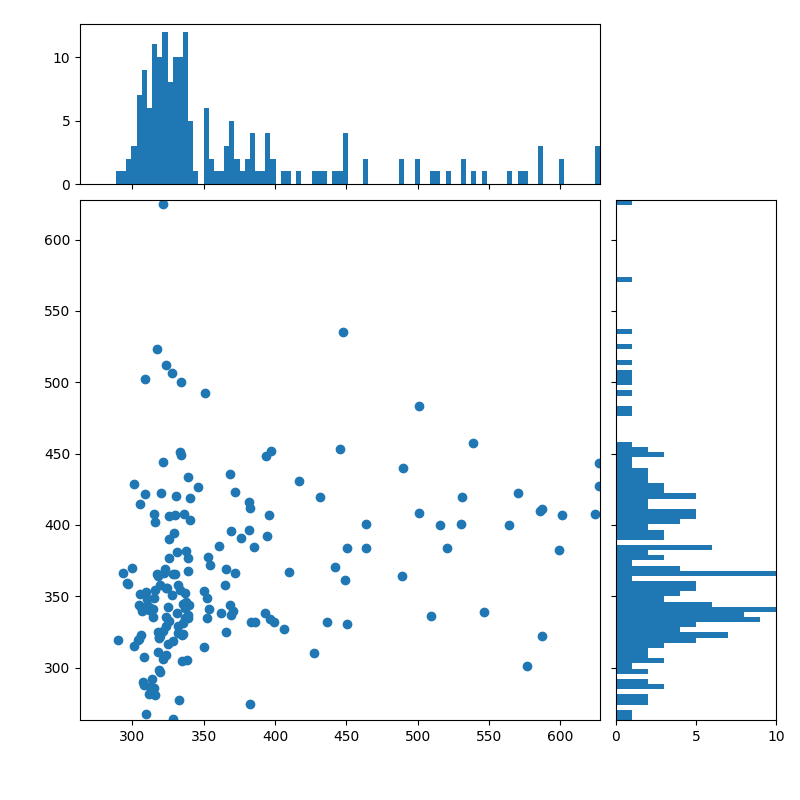}
\hspace{0.5cm}
\includegraphics[width=0.45\textwidth]{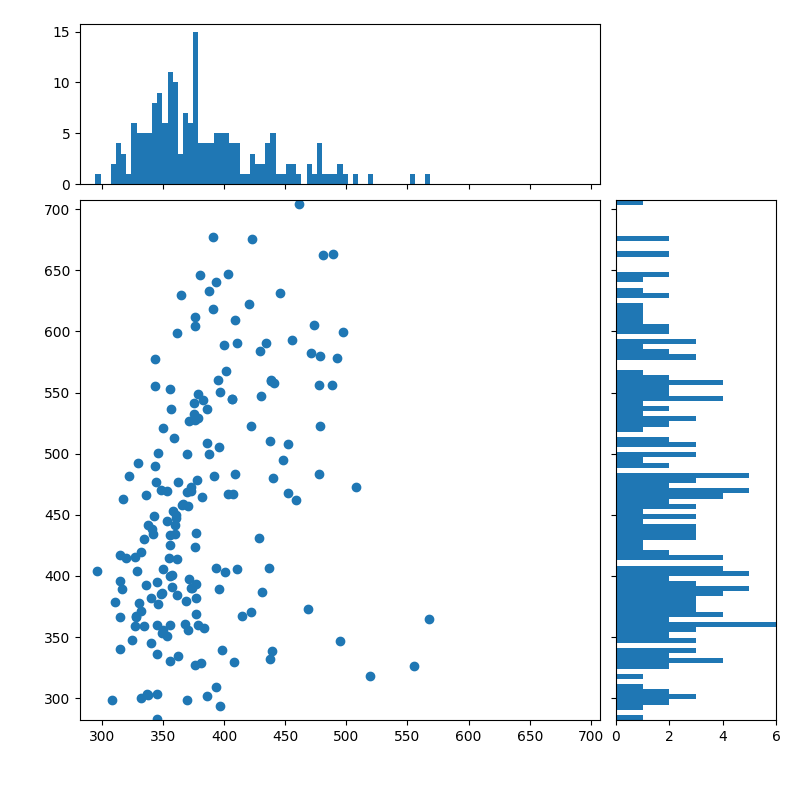}

\includegraphics[width=0.45\textwidth]{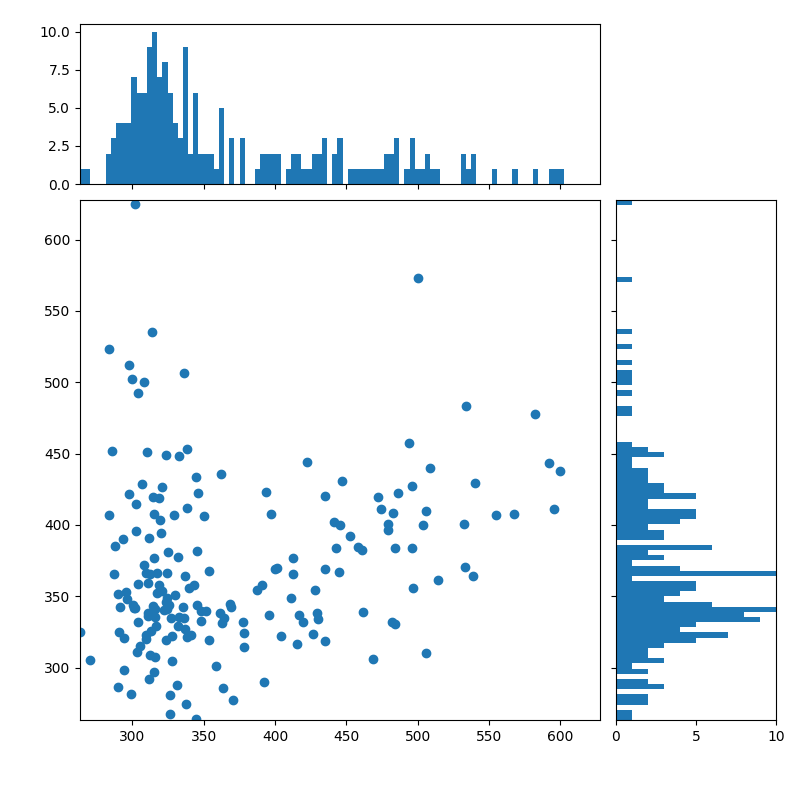}
\hspace{0.5cm}
\includegraphics[width=0.45\textwidth]{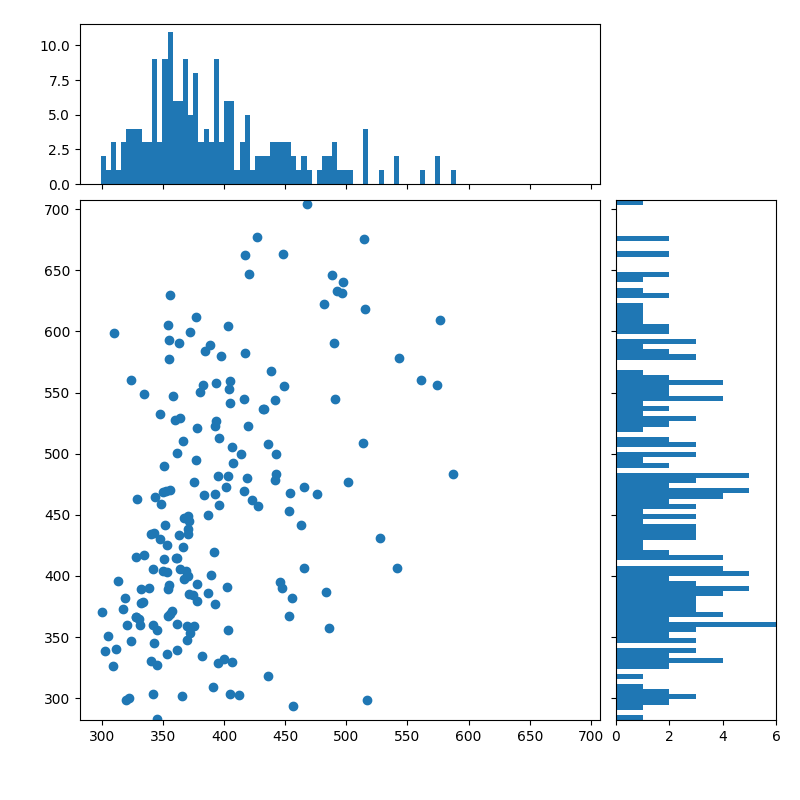}

\includegraphics[width=0.45\textwidth]{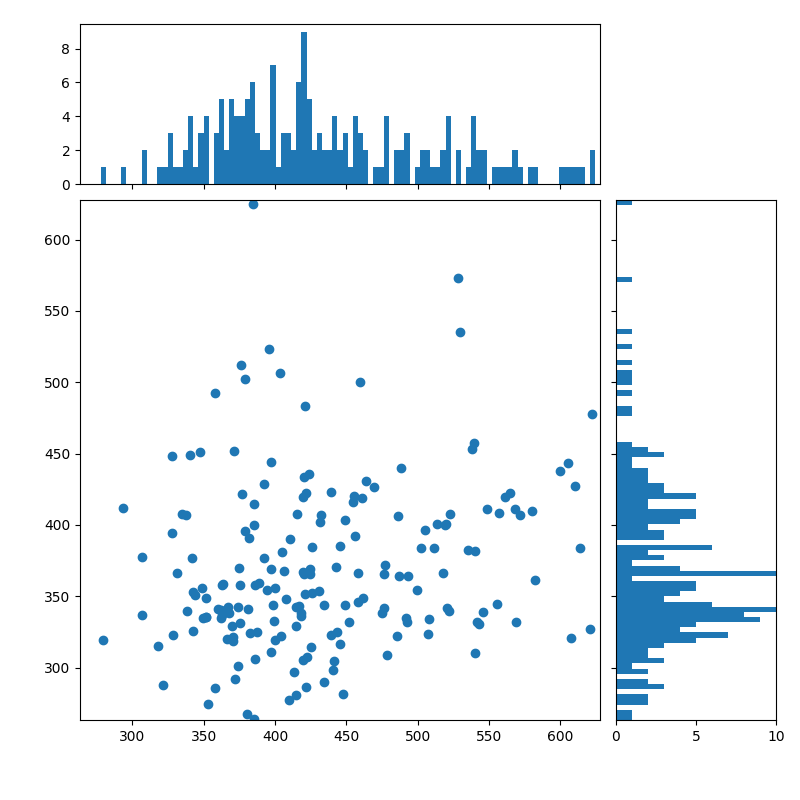}
\hspace{0.5cm}
\includegraphics[width=0.45\textwidth]{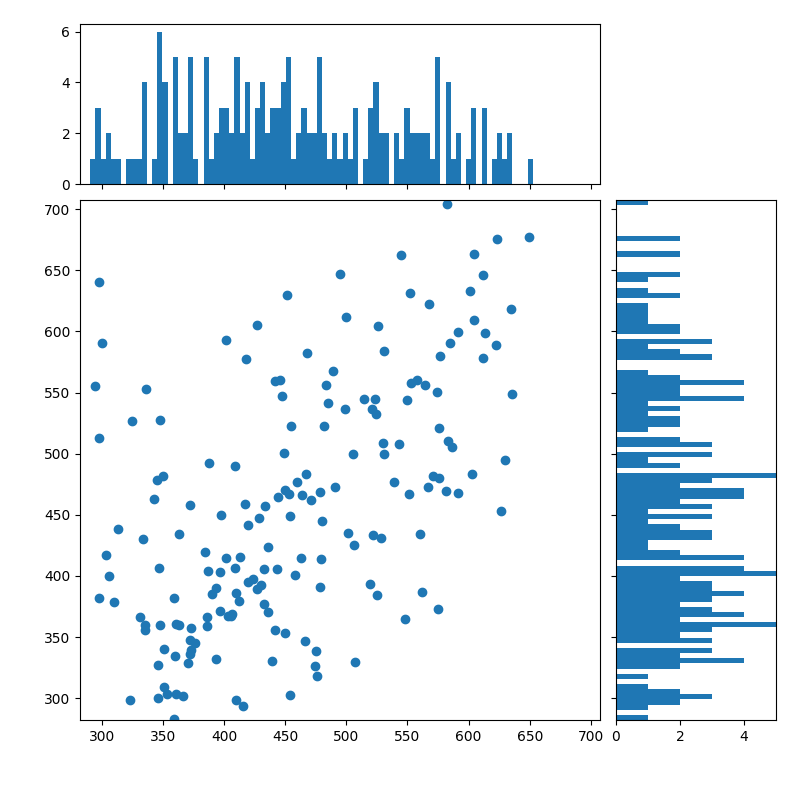}

\caption{Scatter plots and histograms for the wind speed $v_r$ at L1 for 2014 (left) and 2016 (right). Units are $km/s$. The x axis corresponds to the simulation data, the y axis corresponds to the OMNI data. The top row corresponds to the MF model, the middle row to the PFSS model, and the bottom row to the WSA model. \label{fig:scatterv}}
}

\end{figure}

\begin{figure}
{\centering
\includegraphics[width=0.45\textwidth]{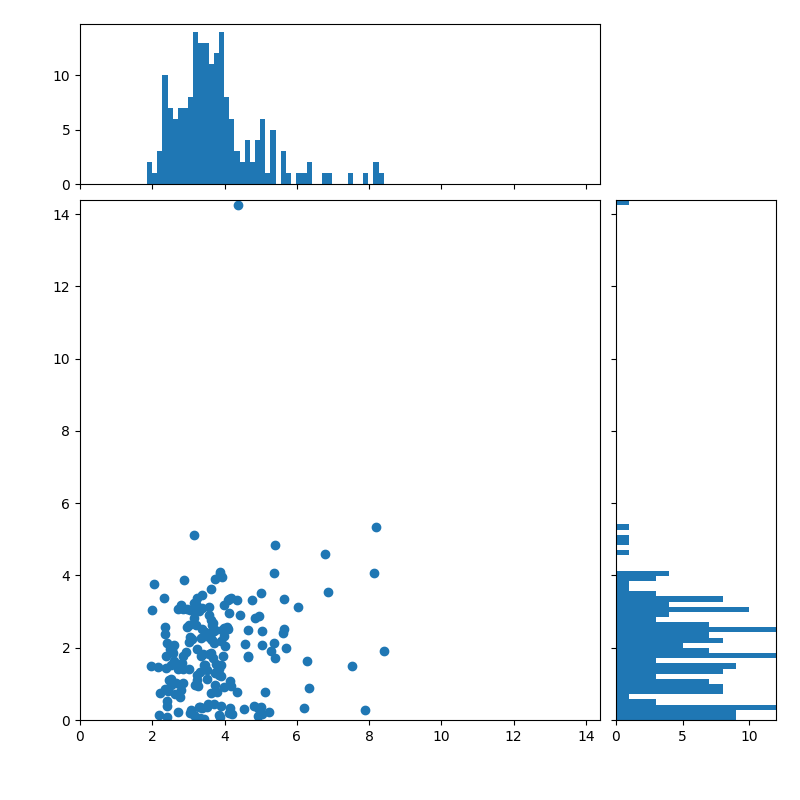}
\hspace{0.5cm}
\includegraphics[width=0.45\textwidth]{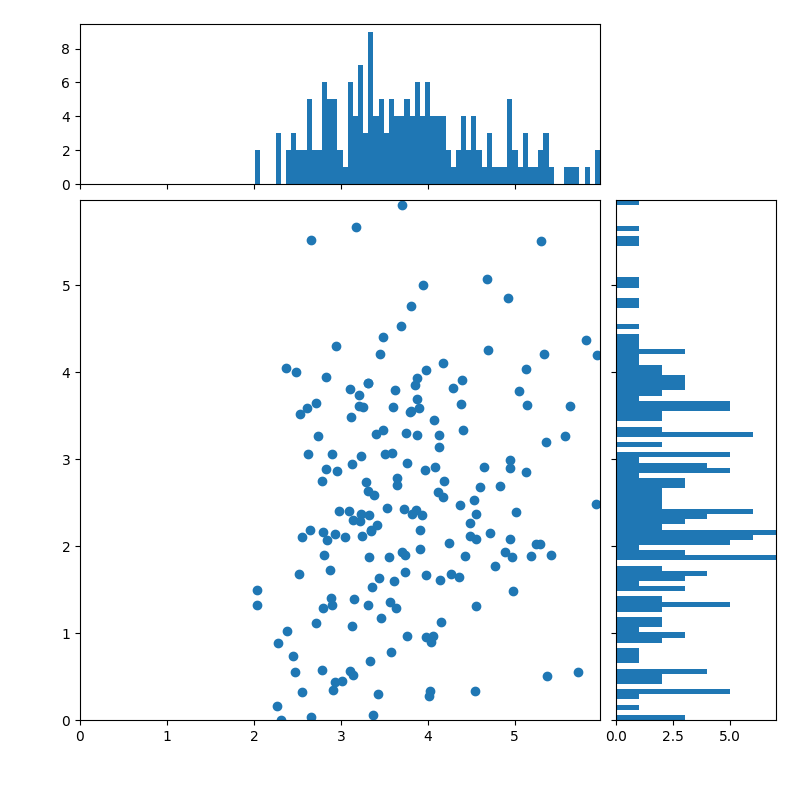}

\includegraphics[width=0.45\textwidth]{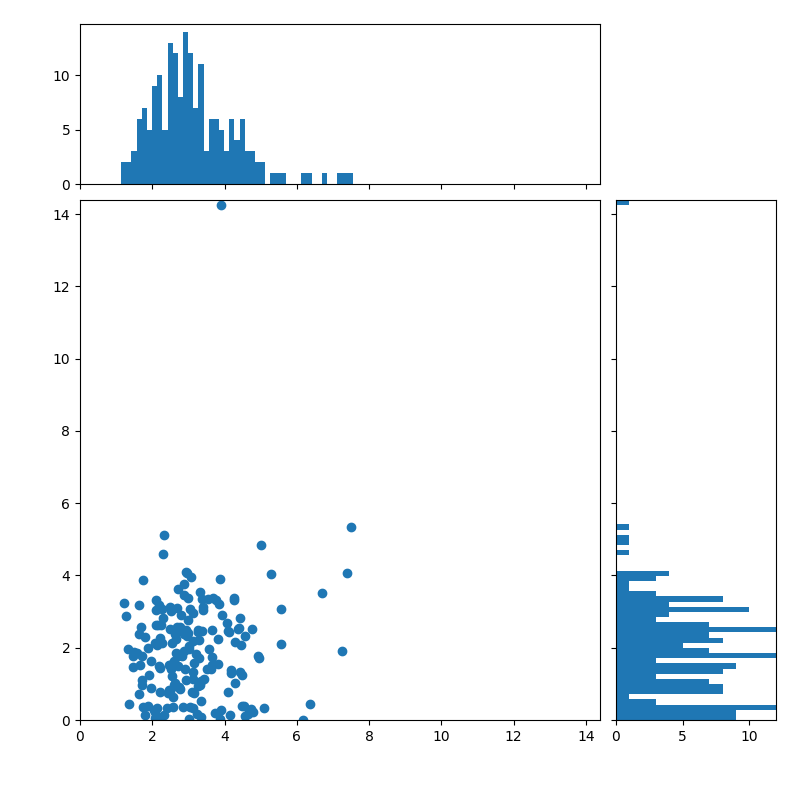}
\hspace{0.5cm}
\includegraphics[width=0.45\textwidth]{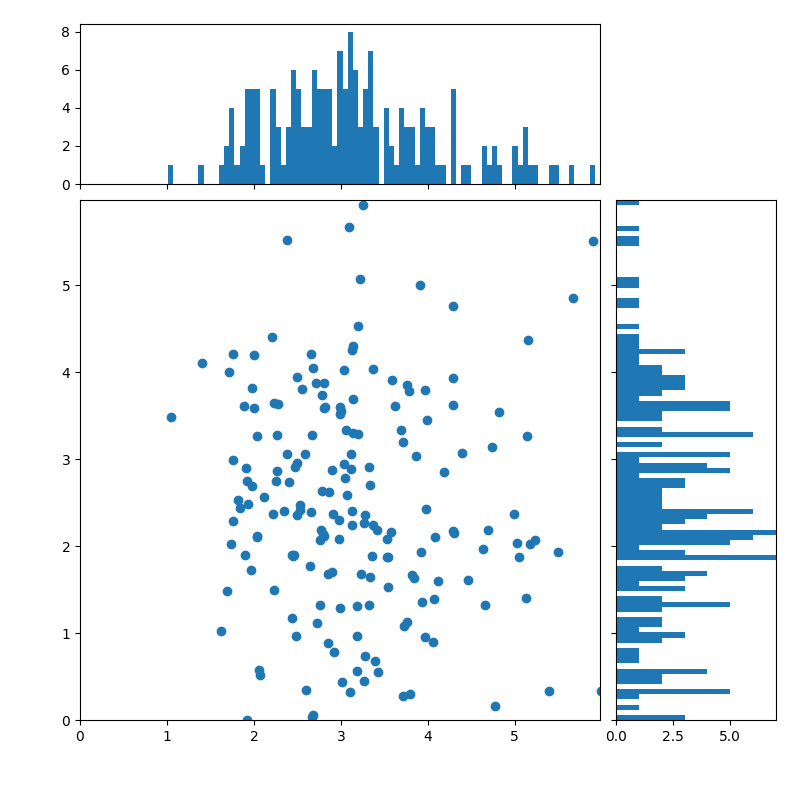}

\includegraphics[width=0.45\textwidth]{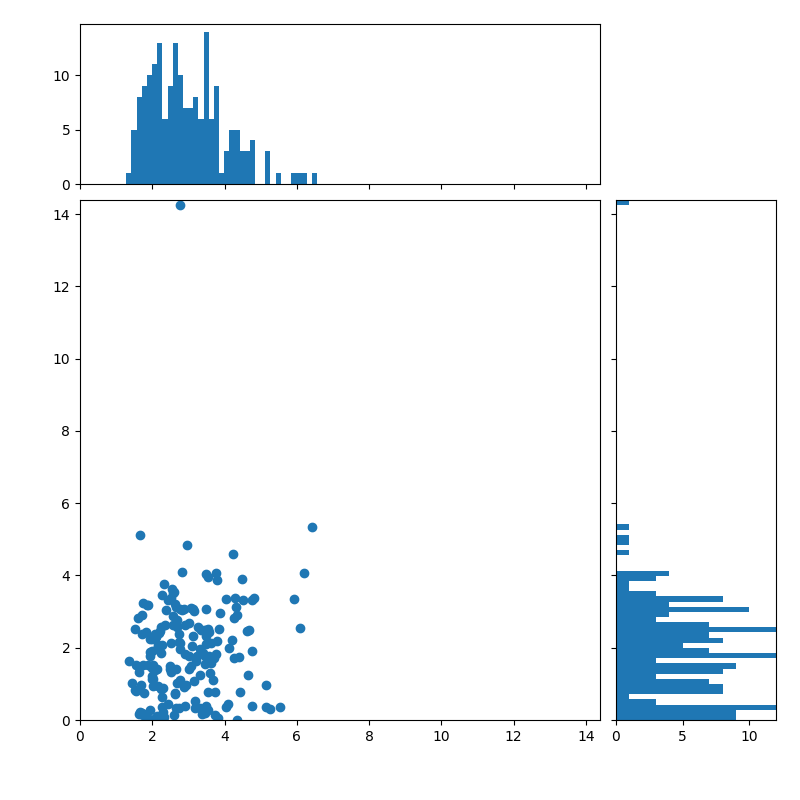}
\hspace{0.5cm}
\includegraphics[width=0.45\textwidth]{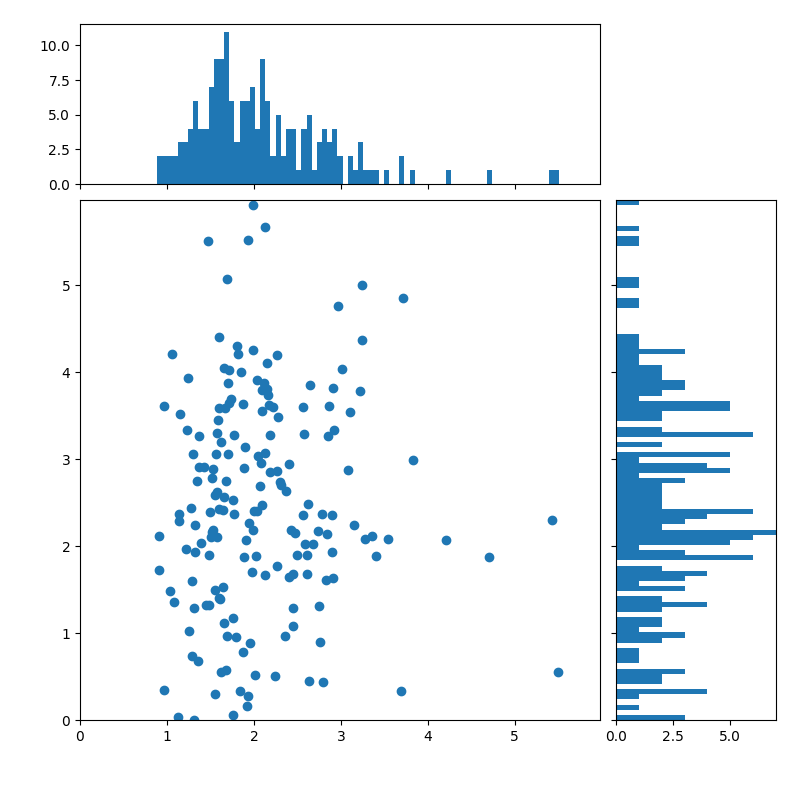}

\caption{Scatter plots and histograms for $B_{mag}$ for 2014 (left) and 2016(right). Units are $nT$. The x axis corresponds to the simulation data, the y axis corresponds to the OMNI data. The top row corresponds to the MF model, the middle row to the PFSS model, and the bottom row to the WSA model.\label{fig:scatterb}}
}
\end{figure}

\begin{figure}
{\centering
\includegraphics[width=0.45\textwidth]{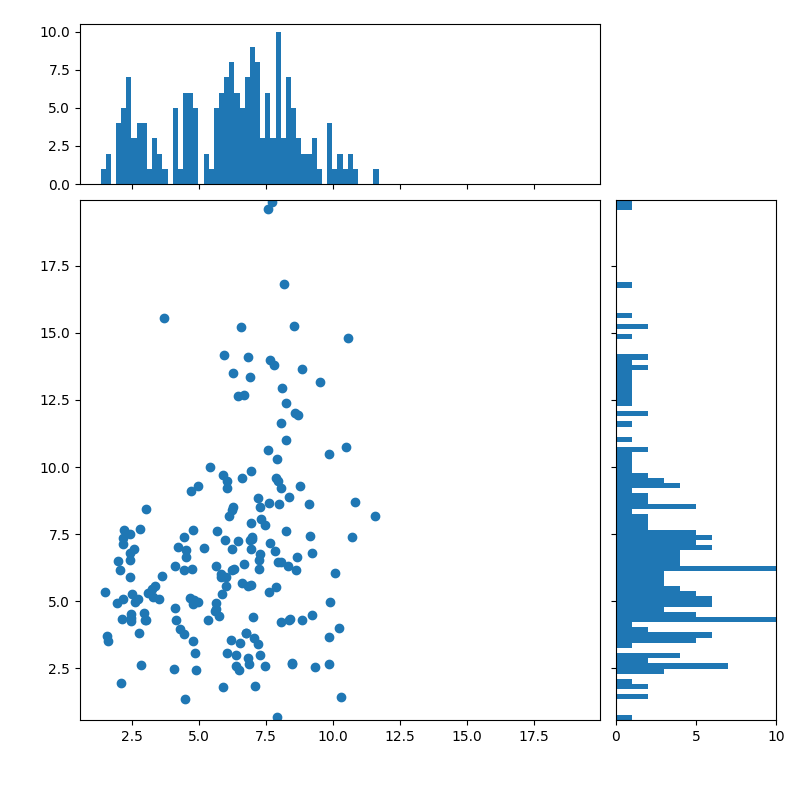}
\hspace{0.5cm}
\includegraphics[width=0.45\textwidth]{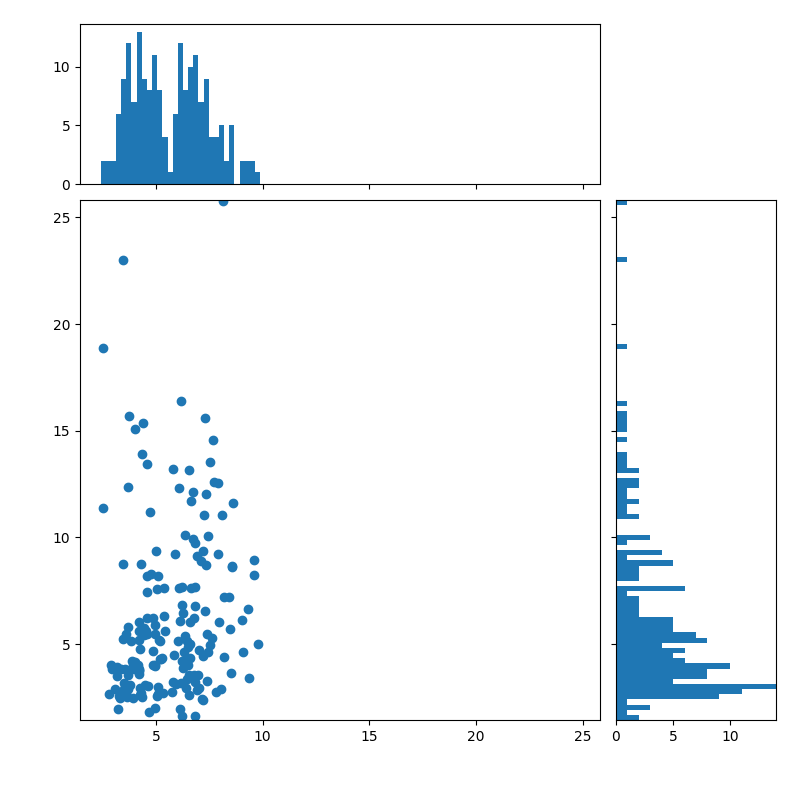}

\includegraphics[width=0.45\textwidth]{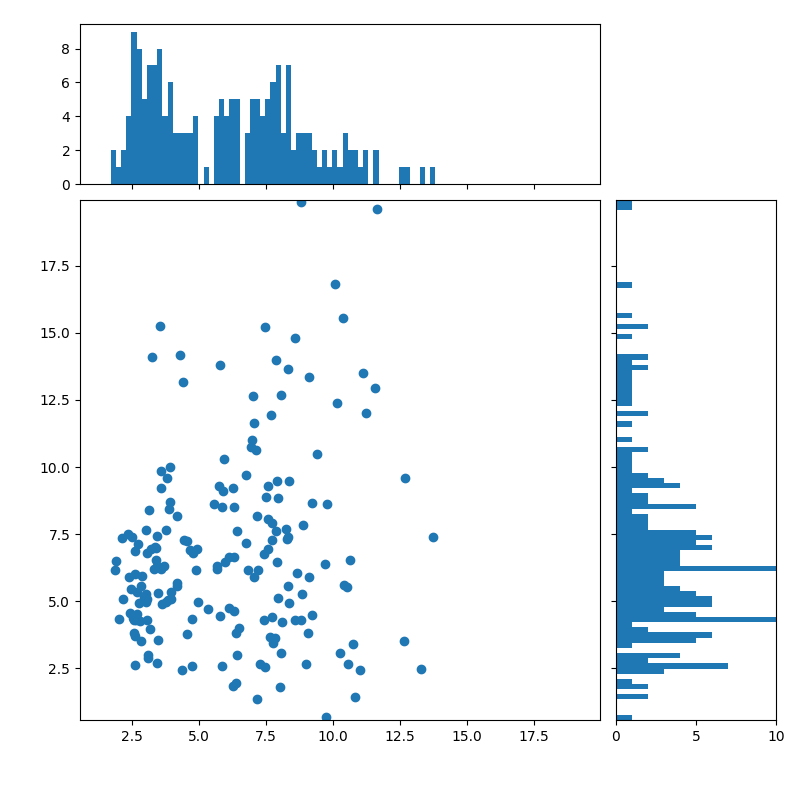}
\hspace{0.5cm}
\includegraphics[width=0.45\textwidth]{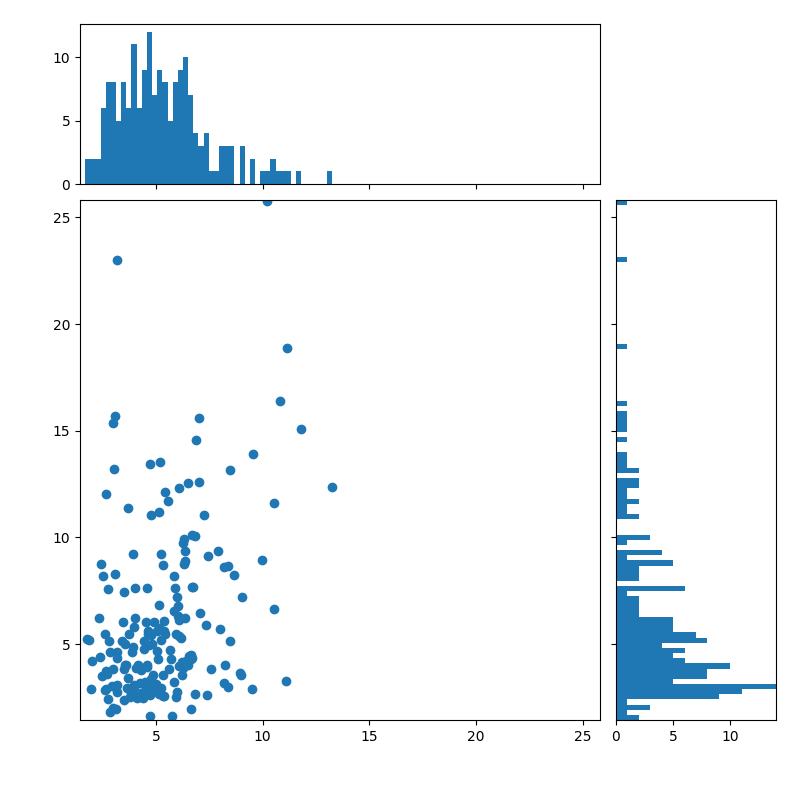}

\includegraphics[width=0.45\textwidth]{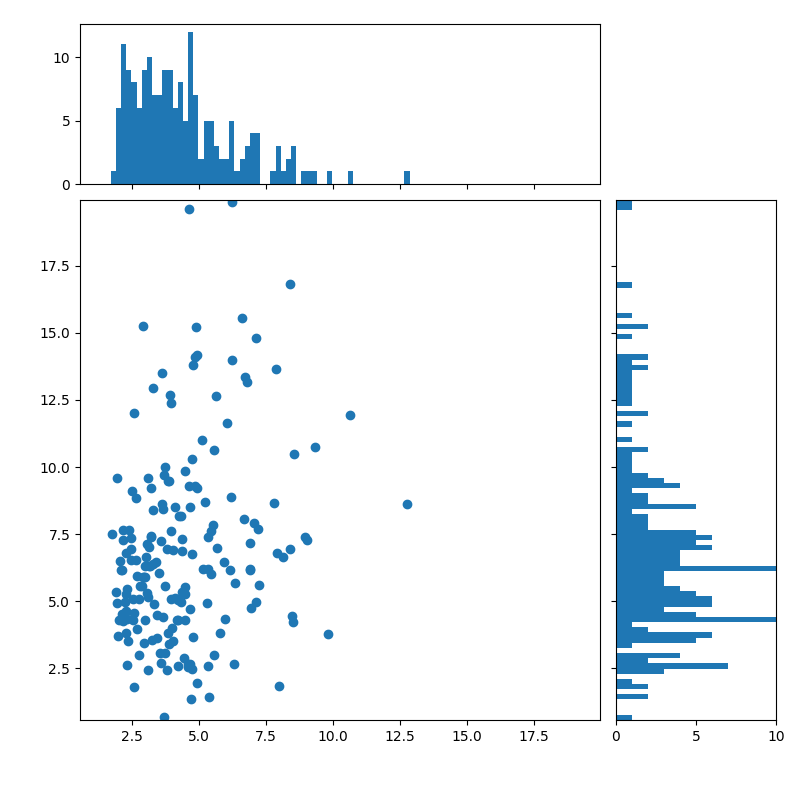}
\hspace{0.5cm}
\includegraphics[width=0.45\textwidth]{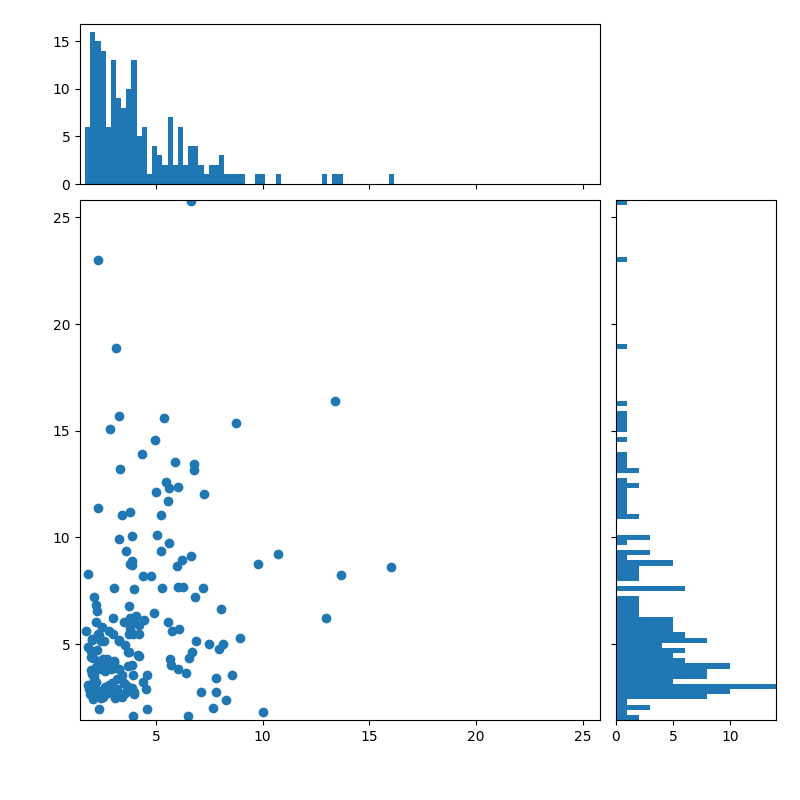}

\caption{Scatter plots and histograms for density for 2014 (left) and 2016 (right). Units are $p/m^{3}$. The x axis corresponds to the simulation data, the y axis corresponds to the OMNI data. The top row corresponds to the MF model, the middle row to the PFSS model, and the bottom row to the WSA model. \label{fig:scatterd}}
}
\end{figure}

\noindent  We first do a comparison in terms of statistical metrics. Figures \ref{fig:scatterv},  \ref{fig:scatterb} and  \ref{fig:scatterd} show scatter plots of the wind speed $v_r$, the magnetic field magnitude $B_{mag}$ and the density $d$, respectively. 
For a quantitative comparison we evaluated the Root Mean Square Error  $RMSE=\sqrt{\frac{\sum_{1}^{n} (f_i-o_i)^2}{n}}$; $bias=\frac{\sum_{1}^{n} (f_i-o_i)}{n}$, and the Pearson correlation coefficient $\rho \in [-1, 1]$ that describes the linear correlation between the simulated and the observed quantities.
Tables \ref{tab:stat_vr},  \ref{tab:stat_bmag} and  \ref{tab:stat_d} summarize the models' performance in these metrics for all variables and both time intervals.

\begin{table}
\caption{Time-series statistics wind speed. The RMSE is in $km/s$. \label{tab:stat_vr}}
\centering
\begin{tabular}{l | r r  }
\hline
 Year (model) & $RMSE$ & $\rho$  \\
\hline

  2014 (MF)  &  94.20 &	0.36   \\
  2014 (PFSS)&  88.57 & 0.20   \\
  2014 (WSA) &  18.15 & 0.19   \\
\hline
  2016 (MF)  & 118.01 & 0.34  \\
  2016 (PFSS)& 111.31 & 0.40  \\
  2016 (WSA) &  92.88 &	0.54  \\
\hline
\multicolumn{3}{l}{}
\end{tabular}
\end{table}

\begin{table}
\caption{Time-series statistics $B_{mag}$. The RMSE is in $nT$. \label{tab:stat_bmag}}
\centering
\begin{tabular}{l | r r  }
\hline
 Year (model) & $RMSE$ & $\rho$ \\
\hline
  2014 (MF)  & 2.54  & 0.16   \\
  2014 (PFSS)& 2.15  & 0.10   \\
  2014 (WSA) & 1.99  & 0.13   \\
\hline
  2016 (MF)  & 1.92 & 0.19    \\
  2016 (PFSS)& 1.90 & -0.11   \\
  2016 (WSA) & 1.56 & -0.006   \\
\hline
\multicolumn{3}{l}{}
\end{tabular}
\end{table}

\begin{table}
\caption{Time-series statistics density. The RMSE is in $p/m^{3}$. \label{tab:stat_d}}
\centering
\begin{tabular}{l | r r }
\hline
 Year (model) & $RMSE$ & $\rho$ \\
\hline
  2014 (MF)  & 3.72 & 0.27   \\
  2014 (PFSS)& 4.12 & 0.19  \\
  2014 (WSA) & 4.32 & 0.24   \\
\hline
  2016 (MF)  & 4.11 & 0.18   \\
  2016 (PFSS)& 3.94 & 0.34   \\
  2016 (WSA) & 4.51 & 0.26  \\
\hline
\multicolumn{3}{l}{}
\end{tabular}
\end{table}

\noindent As we have expected due to the existence of currents in the MF model, this model gives a better correlation with the observed wind speed for solar maximum than the PFSS model, and also than the operational WSA model. Looking at the correlation coefficient for the wind speed and density, the potential 2016 (descending phase) results are better than the 2014 (solar maximum) results. For the magnetic field magnitude, MF performs best for both years. It is interesting to note, however, that this changes when we exclude the time intervals with CMEs from our analysis. In this case, the WSA model performs best in 2016, and the PFSS in 2014 \citep[cf.][]{weinzierl17solarwind}. 
For the density, the MF model again performs better than the potential one in 2014, and not as good in 2016. 

\noindent The RMSE gives a different picture. Here, the two potential models mostly outperform the non-potential one. For the wind speed, the 2014 results have better (lower) values than the 2016 results, while it is the other way around for $B_{mag}$, and mixed for the density. The reason for the poor performance of the MF model can be observed in the scatter plots: The slope of a best-fit line through the data points, is not one. This systematic error could be corrected by optimizing the solar wind speed formula and its parameters. We will briefly discuss such an optimization in Section \ref{sec:conclusion}.

\noindent We conclude that the statistics do not give us the one best method here, and none of the methods are particularly good in terms of real correlation and matching the observations on a one-to-one basis.
However, we can see a tendency that the non-potential model mostly improves the results at solar maximum. 

\subsection{Event-based Validation}\label{sec:val:event}

\noindent In addition to the metrics defined in the previous section, event-based validation is crucial in assessing the various models. For the purpose of this study, the focus is on the arrival time of slow-to-fast stream interaction regions (SIRs). The error on the arrival time of SIRs is typically of the order of one day, and so larger than the error on arrival time of CMEs. The arrival of SIRs can trigger a geomagnetic storm response and the fast solar wind is associated with high electron fluences in the radiation belts. One of the drivers of developing the MF model was to get a better characterization of coronal holes, associated with the fast solar wind.

\begin{figure}
\includegraphics[width=\textwidth]{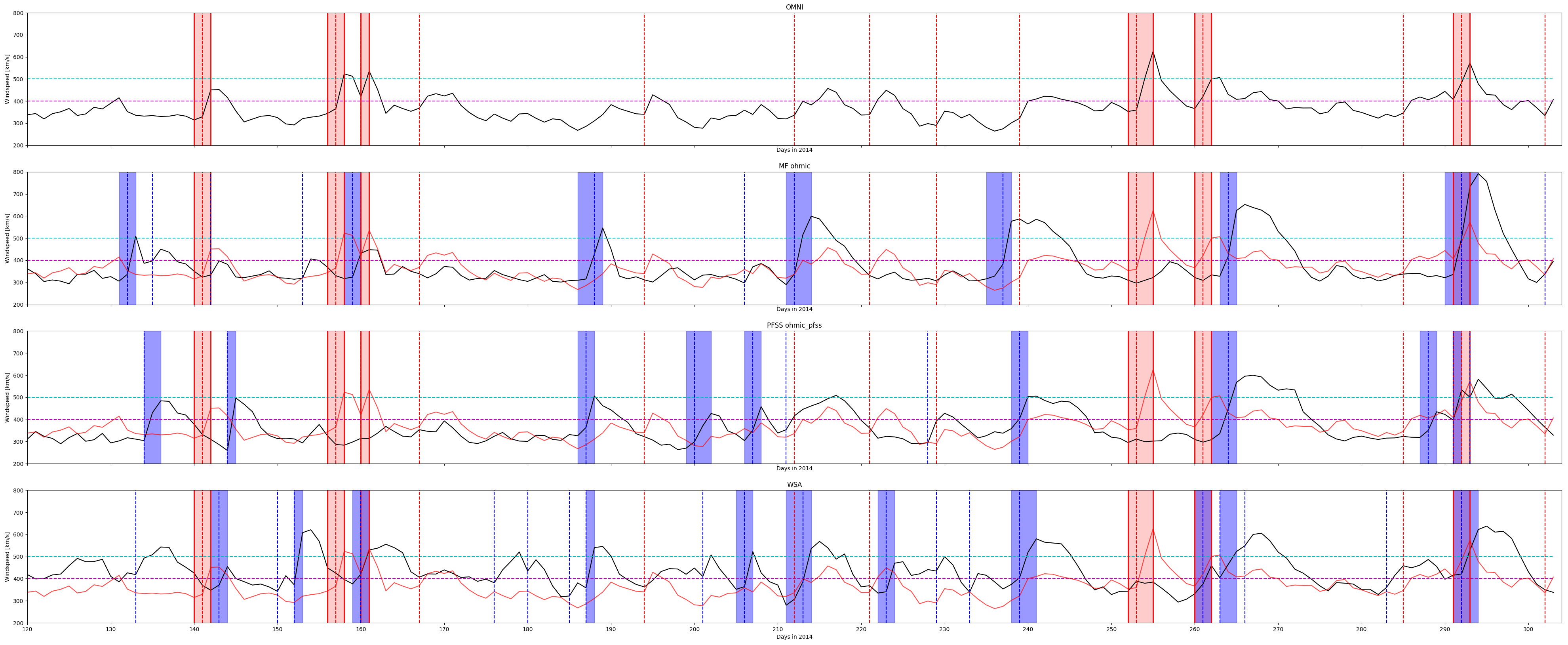}\\

\includegraphics[width=\textwidth]{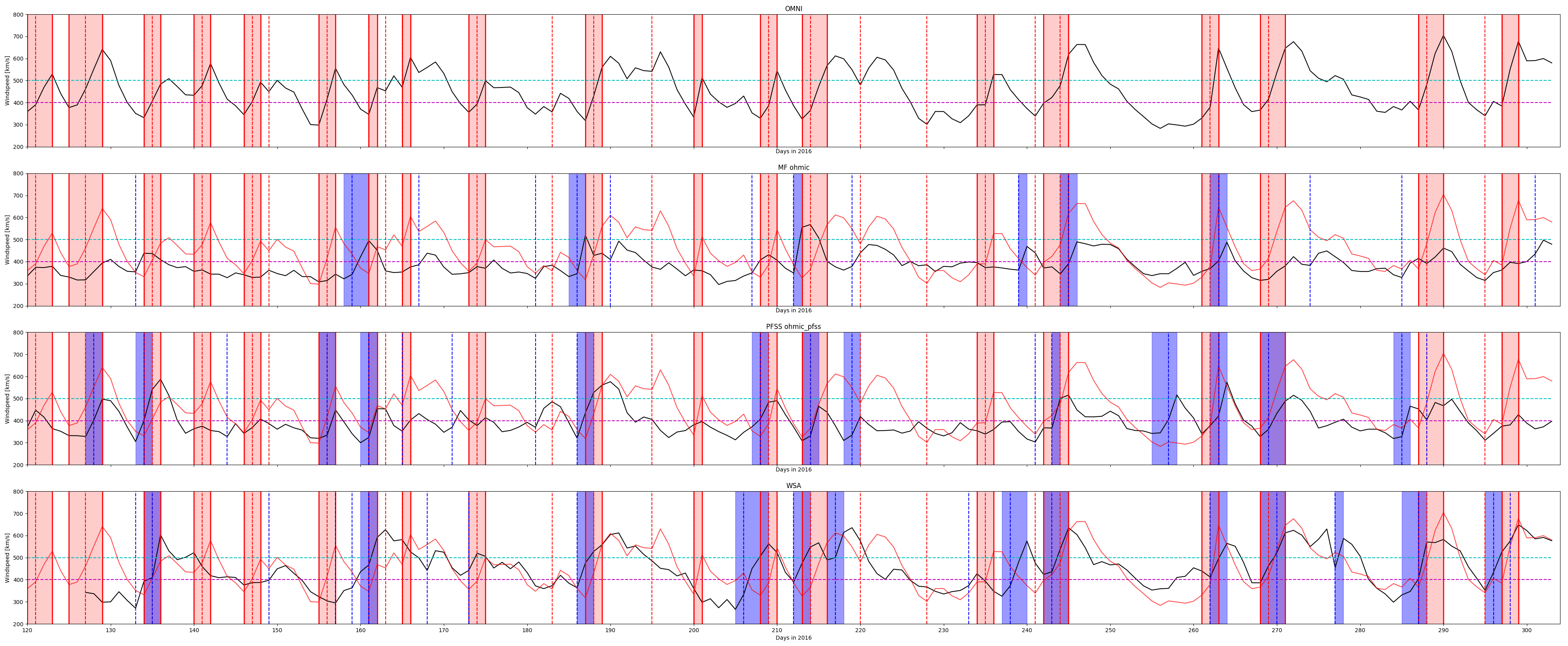}
\caption{SIR detection results for 2014 (top) and 2016 (bottom). Detected SIRs from OMNI observations (top row) are marked in red and and copied down to the other rows (second row is MF, third row is PFSS, bottom row WSA) for easier comparison, as well as the velocity profile (red line in simulation plots). Detected events for the simulations are highlighted in blue. Dashed vertical red (OMNI) and blue (simulations) lines mark detected HSEs.  \label{fig:SIR} }
\end{figure}

\noindent The SIR detection algorithm was originally developed in \citep{owens2005eventbased} and further refined in \citep{macneice2009events, jian2015ccmcmodelvalidation, jian2016ccmcmodelvalidation}. We use here, for comparability, the thresholds as given in \citep{jian2015ccmcmodelvalidation}. In Figure \ref{fig:SIR} we show how our wind models perform in terms of predicting SIRs and high speed enhancements (HSEs), i.e., abrupt accelerations in $v_r$.

\noindent There exists a great number of metrics, or skill scores, for assessing the ``goodness'' of a binary forecast (cf., for example, \citep{barnes2016flareforecast}). They are usually based on the number $n_{hit}$ of events that very correctly predicted, the number $n_{miss}$ of events that were missed in the prediction and the number $n_{false}$ of false alarms. 
One such skill score, which we will use here, is the critical success index or threat score (TS, see, e.g., \citep{schaefer90criticalsuccessindex} and references therein) $TS=\frac{n_{hit}}{n_{hit}+n_{miss}+n_{false}}$. The TS assumes that it is most important to avoid missing an event (i.e., keep $n_{miss}$ low), even on the cost of increasing the number of false alarms.

\begin{table}
\caption{SIR detection performance. \label{tab:SIR}}
\centering
\begin{tabular}{l | c c c c c}
\hline
 Year (model) & $n_{hit}$ & $n_{miss}$ & $n_{false}$ & TS\\
\hline
  2014 (MF)  & 3 & 3 & 4 & 0.3 \\
  2014 (PFSS)& 3 & 3 & 5 & 0.27 \\
  2014 (WSA) & 5 & 1 & 6 & 0.4 \\
  \hline
  2016 (MF)  & 6 & 14 & 0 & 0.3 \\
  2016 (PFSS)& 11 & 8 & 1 & 0.55 \\
  2016 (WSA) & 11 & 7 & 2 & 0.55 \\
\hline
\multicolumn{5}{l}{}
\end{tabular}
\end{table}

\noindent Table \ref{tab:SIR} summarizes the SIR prediction results. Overall, the WSA model performs best in predicting the SIRs correctly. It has the highest TS for both time intervals, with PFSS performing equally well in this metric for the descending phase. The difference is that PFSS has one more miss and one more false alarm in the 2016 interval. The MF model has no false alarms for this case, but a lot of misses.
For solar maximum, the result is not as clear. The MF model performs slightly better than the PFSS model in terms of TS and false alarms. The WSA model has a better TS and a higher hit rate, but also more false alarms. 

\noindent Looking at the velocity profile in Figure \ref{fig:SIR}, we can see that for some of the missed SIRs in the MF and PFSS model, there actually is a peak around that time, only not as high as required for the SIR detection algorithm. This, as in the scatter plots, points to a systematic error. We conclude that a better parameter optimization for the solar wind speed formula might improve the performance here. In addition, an adaption of the SIR detection thresholds would lead to more hits, but, as mentioned before, we have decided to leave the thresholds as in \citep{jian2015ccmcmodelvalidation} for better comparability.


\section{Conclusion and Future Work \label{sec:conclusion}}

\noindent We have compared the performance of three different simulation pipelines for solar wind prediction, using GONG or ADAPT maps as input, non-potential or potential models for the solar corona, and Enlil for extrapolating the results to 1AU. These models were compared  with OMNI observations of wind speed, magnetic field strength, and density. We did both a statistical analysis and an evaluation of their performance in predicting SIRs; and we considered two six-months periods, one at solar maximum and one in the descending phase of the solar cycle. 

\noindent We observed a difference in performance of the models due to the phase in the solar cycle. At solar maximum, the non-potential model performed best in the statistical metrics, while the potential and WSA model were better in the descending phase. The better performance of the MF model during solar maximum was expected, as this model, in contrast to the potential ones, includes currents in the corona, which are more frequent at this time. In the event-based validation, the operational WSA model always performed best in terms of hits and threat score, but it also produced the most false alarms. 

\noindent Future work will include testing the effect of re-introducing the expansion factor term (in adapted form) into the empirical wind speed formula for the DuMFriC potential and non-potential simulations. The models then have to be tested with different input data, i.e., from different observatories and different forms of synoptic maps, and extended from simulating the purely ambient wind to including CMEs. The usage of a local method for computing the electric field, avoiding the ``halos'' as described in \citet{weinzierl16photosphericdriving}, would be expected to improve the MF model \citep[cf.][]{yeates17sparseefield}.

\noindent The next crucial step will be to use automatic optimization methods for parameter fitting. Although doing this was out of scope of this paper, we give here some considerations concerning automatic parameter optimization.

\noindent It is generally assumed that the major part of the wind speed evolution happens close to the Sun, i.e., between the Sun and 0.3 AU \citep{rosenbauer77evolution, schwenn78evolution}, although \citet{mcgregor2011windevolution} found that, especially at solar minimum and for intermediate wind speeds, there is still some significant change in the wind speed between 0.385 AU and 1 AU. An optimization using simulation data at 1 AU, however, is very time consuming as it requires running an MHD simulation as e.g. Enlil or an appropriate approximation (see \citep{riley2011mappingfromsunto1au} for a comparison of techniques). Therefore, as a first attempt, we could try and compare the speed distributions of the simulation at 0.1 AU and observations at 1 AU for receiving a parameter fit. We expect this to yield better results during solar maximum than during solar minimum.

\noindent In order to evaluate a suitable method for fitting the parameters there are a number of metrics for the distribution of the solar wind speed values that could be considered. The metrics we propose to consider are the root mean squared error (RMSE), the correlation factor, and the shape of the histogram, i.e. the difference of the mean values, the difference of the variance and the difference of the skewness.

\noindent Also, a simultaneous (multi-objective) optimization of a subset of these metric and optimization of single objectives should be tested. A multi-objective optimization determines the Pareto front of multiple fitness functions, that is, the points where improving the score for one objective would worsen the score for the other. Here, another question to consider is how to choose the "best" point of the resulting set: by minimizing one of the objectives, or by using some kind of trade-off? There exist a number of sophisticated methods alone for the purpose of choosing the the optimal point on the Pareto front. 

\noindent We have seen in Section \ref{sec:results}, though, that all these metrics are only of limited value for parameter optimization for operational space weather forecasting. As noted by other authors \citet{owens2005eventbased, macneice2009events} before, an optimal correlation factor or RMSE is no guarantee for a good forecast w.r.t. relevant space weather events. Therefore, parameter fitting should ultimately be based on the ability to predict space weather events, such as the arrival time of stream interaction regions and high speed enhancements (see Section \ref{sec:val:event}). This, however, is computationally very expensive, as it requires the whole simulation pipeline to be run a lot of times. We leave this as an interesting direction for future research which has to combine mathematical, physical and computer science skills.

\section{Acknowledgments}
\noindent The work utilizes data produced collaboratively between Air Force 
 Research Laboratory (AFRL) and the National Solar Observatory (NSO). The 
 ADAPT model development is supported by AFRL and AFOSR. The input data utilized by ADAPT are obtained by NSO/NISP (NSO Integrated Synoptic Program). NSO is operated by the Association of Universities for Research in Astronomy (AURA), Inc., under 
 a cooperative agreement with the National Science Foundation (NSF). The OMNI data were obtained from the GSFC/SPDF OMNIWeb interface at http://omniweb.gsfc.nasa.gov.
MW and ARY thank STFC and the AFOSR for financial support.

\end{document}